\documentclass[aps,prb,a4paper,twocolumn,showpacs,showkeys,fleqn,floatfix,unsortedaddress]{revtex4-1}

\usepackage{graphicx}
\usepackage{color}
\usepackage{amsmath}
\usepackage{multirow}

\newcommand{\figwidth}{\columnwidth}

\newcommand{\vect}[1]{\mathbf{#1}}

\begin{document}

\title{Reminiscence of a magnetization plateau in a magnetization processes of toy-model triangular and tetrahedral clusters.}

\author{V.N.Glazkov}
\email{glazkov@kapitza.ras.ru}
\affiliation{Kapitza Institute for Physical Problems RAS, Kosygin str. 2, 119334 Moscow, Russia}

\affiliation{International Laboratory for Condensed Matter Physics, National research university ``Higher School of Economics'', Myasnitskaya str. 20, 101000 Moscow, Russia}

\begin{abstract}
 We discuss magnetization curves of a toy-model trigonal and tetrahedral clusters. Nonlinearity of magnetization with local minimum of differential susceptibility resembling known magnetization plateaus of triangular-lattice and pyrochlore lattice antiferromagnets is observed at intermediate temperature range $ J\lesssim T\lesssim\Theta$ (here $J$ is the exchange coupling constant and $\Theta$ is a Curie-Weiss temperature). This behavior is due to increased statistical weight of the states with intermediate total spin of the cluster, which is related to the ``order-by-disorder'' mechanism of plateau stabilization of a macroscopic frustrated magnet.
 \end{abstract}

\maketitle

\section{Introduction}

Triangular motives in a crystallographic lattice of a magnet frequently results in the frustration of exchange bonds which leads to unconventional form of magnetic ordering or even to the formation of the disordered spin-liquid state, antiferromagnets on the triangular, kagom\'{e} or pyrochlore lattice are the well studied examples of this effect \cite{ramirez,gardner,lacroix}. Strong degeneracy of the ground state of the frustrated magnet makes thermal and quantum fluctuations  (``order-by-disorder'' mechanism) important when selecting a ground state of the magnet. This produces remarkable effect on the low-temperature magnetization curves: Heisenberg antiferromagnet on a triangular lattice demonstrates magnetization plateau at 1/3 of  saturation magnetization value \cite{chubukov}, similar plateau at 1/3 of saturation value is predicted for kagom\'{e} lattice Heisenberg antiferromagnet \cite{nishimoto}, Heisenberg antiferromagnet on a pyrochlore lattice demonstrates a magnetization plateau at 1/2 of saturation magnetization,\cite{motome} some other models also demonstrate magnetization plateaus \cite{honeker,morita}. Magnetization plateaus of triangular (and kagom\'{e})  and pyrochlore lattice antiferromagnets  correspond to the collinear phases ($uud$ and $uuud$, correspondingly), which are stabilized by ``order by disorder'' mechanism. These plateaus  were considered in details theoretically \cite{lacroix,chubukov,motome} and observed experimentally in a variety of magnets \cite{lacroix,rbfeplateau,rbfesmirnov,spinelplateau1,spinelplateau2,okuta}.

Thus, stabilization of the collinear ordered state  in a wide field interval, unexpected in a mean field approximation,   is a well known feature of triangular lattice antiferromagnets and of some other systems. This effect is accompanied by magnetization plateau which is frequently observed as a wide range of essential diminishing  of differential susceptibility $\chi=\partial M/\partial B$.  Present work was stimulated by the observations of Refs.\cite{rbfesmirnov,rbfediluted} which demonstrated that minimum of differential susceptibility is observed in triangular lattice antiferromagnet {\em above} the Neel temperature\cite{rbfesmirnov} as well as it is observed in a triangular lattice antiferromagnet  with a chaotic modulation of the exchange bonds via the substitiution of a portion of  nonmagnetic ions on a crystal lattice site  by other nonmagnetic ions, where collinear phase is suppressed by exchange couplings randomness \cite{rbfediluted}.

In the present report we describe results of the analysis of a related toy models, which are the building block of aforementioned frustarted lattices: equilateral triangular and tetrahedral clusters. We will demonstrate that, quite unexpectedly, magnetization curves of these compact models contain reminiscence of the magnetization plateaus visible at the temperatures of the order of Curie-Weiss temperature: differential susceptibility has a local minimum at net magnetization values close to those of plateaus in macroscopic magnets.This effect is present both for quantum and classical spins, its origin is purely statistical and is microscopic scale analogue of a macroscopic ``order by disorder'' mechanism: it is due to the higher weight of the cluster states with intermediate total spins. We present detailed analysis of these toy models and compare model predictions with the aforementioned known results\cite{rbfesmirnov,rbfediluted} for triangular lattice antiferromagnet RbFe(MoO$_4$)$_2$.

\section{Toy model formulation}

We will consider clusters of $N=3$ and 4 spins $S$ with each spin equally antiferromagnetically coupled to all other spins. This corresponds to the triangular and tetrahedral geometry of the exchange bonds and will be referred as triangular (trigonal) and tetrahedral clusters further on. Hamiltonian of quantum model (here $\left\langle i,j \right\rangle$ means that each pair is counted only once)

\begin{equation}\label{eqn:quantum_model}
    \hat{\cal H}=J \sum_{\left\langle i,j \right\rangle} \hat{\vect{S}}_i\hat{\vect{S}}_j-g\mu_B B \sum_i \hat{S}_{z,i}
\end{equation}

\noindent
can be rewritten in the terms of the total spin $S_{tot}$

\begin{equation}\label{eqn:quantum_rewritten}
    \hat{\cal H}=\frac{J}{2}\hat{\vect{S}}_{tot}^2-g\mu_B B \hat{S}_{z,tot}-\frac{N}{2} J S (S+1)
\end{equation}

\noindent
since $\vect{S}_{tot}^2=\sum \vect{S}_i^2+2 \sum_{\left\langle i,j \right\rangle} \vect{S}_i\vect{S}_j$. Last additive term in Eqn.(\ref{eqn:quantum_rewritten}) is a matter of zero energy choice and we will omit it later on.

Similarly one can formulate classical model with unit vectors $\vect{S}$ in the vertices of a triangle or a tetrahedron. Full energy of such a cluster can be similarly expressed through the total spin

\begin{eqnarray}
    E&=&J \sum_{\left\langle i,j \right\rangle} {\vect{S}}_i{\vect{S}}_j-g\mu_B B \sum_i {S}_{z,i}=\nonumber\\
    &=&\frac{J}{2}\vect{S}_{tot}^2-g\mu_B B S_{z,tot}-\frac{N}{2} J S^2
\label{eqn:classic model}
\end{eqnarray}

\noindent
here, again, last additive term is simply a matter of zero energy choice and will be omitted in calculations.

We are interested here in the magnetization $m_z(B,T)$ and in the differential susceptibility $\chi(B,T)=\frac{\partial m_z}{\partial B}$, which are the experimentally measurable quantities. They can be straightforwardly calculated via standard thermodynamic averaging: $Z=\sum_{\left| ... \right\rangle}e^{-E/T}$, here summation runs over all possible eigenstates of Hamiltonian, $F=- T \ln Z$, $M=-\partial F/\partial B$ etc.

Because of the high symmetry of the clusters under consideration energy depends only on total spin and its projection. Thus summation over all states can be replaced by the summation over possible total spin values $S_{tot}$ (which runs from 0 or 1/2 to $N S$) weighted with weight factor $D_N(S_{tot})$ and its projections. The weight factor is the number of combinations yielding particular value of total spin for quantum model and density of such combinations for classical model. This yields

\begin{eqnarray}
Z&=&\sum_{S_{tot}} D_N(S_{tot}) \sum_{S_z} e^{-E/T} \label{eqn:Z-quant}\\
Z&=&\int_0^{N S} D_N(S_{tot}) \int_0^\pi 2\pi sin\Theta e^{-E/T} dS d\Theta=\nonumber\\
&=&4\pi \int_0^N D_N(S) \frac{sh\left(\frac{g\mu_B BS}{T}\right)}{\frac{g\mu_B BS}{T}} e^{-\frac{J S^2}{2T}} dS \label{eqn:Z-classic}
\end{eqnarray}

\noindent
for quantum and classical case correspondingly.

All analysis of the magnetization and differential susceptibility was done using GNU Octave \cite{octave} software and default utilities for integration and minimization, whenever necessary.

\section{Calculated weight factors}

We have calculated weight factors for quantum model and tabulated it in the Tables \ref{tab:D3even}, \ref{tab:D3odd} and \ref{tab:D4}. We did not obtained compact expressions for the weight factors of quantum model.

\begin{table}
\caption{Weight factors $D_3(S_{tot})$ for triangular ($N=3$) clusters with integer spin}
\label{tab:D3even}
\begin{tabular}{c|ccccccccccccccccc}
  \hline
  S & \multicolumn{16}{|c}{Total spin $S_{tot}$} \\
  \cline{2-17}
   &0 & 1 & 2 & 3 & 4 & 5 & 6 & 7 & 8 & 9 & 10 & 11 & 12 & 13 & 14 & 15 \\
  \hline
  1 & 1 & 3 & 2 & 1 & --- & --- & --- & --- & --- & --- & --- & --- & --- & --- & --- & --- \\
  2 & 1 & 3 & 5 & 4 & 3 & 2 & 1 & --- & --- & --- & --- & --- & --- & --- & --- & --- \\
  3 & 1 & 3 & 5 & 7 & 6 & 5 & 4 & 3 & 2 & 1 & --- & --- & --- & --- & --- & --- \\
  4 & 1 & 3 & 5 & 7 & 9 & 8 & 7 & 6 & 5 & 4 & 3 & 2 & 1 & --- & --- & --- \\
  5 & 1 & 3 & 5 & 7 & 9 & 11 & 10 & 9 & 8 & 7 & 6 & 5 & 4 & 3 & 2 & 1 \\
  \hline
\end{tabular}
\end{table}

\begin{table*}
\caption{Weight factors $D_3(S_{tot})$ for triangular ($N=3$) clusters with half-integer spin}
\label{tab:D3odd}
\begin{tabular}{c|cccccccccccccc}
  \hline
  S & \multicolumn{14}{|c}{Total spin $S_{tot}$}\\
  \cline{2-15}
   & 1/2 & 3/2 & 5/2 & 7/2 & 9/2 & 11/2 & 13/2 & 15/2 & 17/2 & 19/2 & 21/2 & 23/2 & 25/2 & 27/2 \\
  \hline
  1/2 & 2 & 1 & --- & --- & --- & --- & --- & --- & --- & --- & --- & --- & --- & --- \\
  3/2 & 2 & 4 & 3 & 2 & 1 & --- & --- & --- & --- & --- & --- & --- & --- & --- \\
  5/2 & 2 & 4 & 6 & 5 & 4 & 3 & 2 & 1 & --- & --- & --- & --- & --- & --- \\
  7/2 & 2 & 4 & 6 & 8 & 7 & 6 & 5 & 4 & 3 & 2 & 1 & --- & --- & --- \\
  9/2 & 2 & 4 & 6 & 8 & 10 & 9 & 8 & 7 & 6 & 5 & 4 & 3 & 2 & 1 \\
  \hline
\end{tabular}
\end{table*}

\begin{table*}
\caption{Weight factors $D_4(S_{tot})$ for tetrahedral ($N=4$) clusters}
\label{tab:D4}
\begin{tabular}{c|ccccccccccccccccccccc}
  \hline
  S & \multicolumn{21}{|c}{Total spin $S_{tot}$}\\
  \cline{2-22}
   & 0 & 1 & 2 & 3 & 4 & 5 & 6 & 7 & 8 & 9 & 10 & 11 & 12 & 13 & 14 & 15 & 16 & 17 & 18 & 19 & 20 \\
  \hline
  1/2 & 2 & 3 & 1 & --- & --- & --- & --- & --- & --- & --- & --- & --- & --- & --- & --- & --- & --- & --- & --- & --- & --- \\
  1 & 3 & 6 & 6 & 3 & 1 & --- & --- & --- & --- & --- & --- & --- & --- & --- & --- & --- & --- & --- & --- & --- & --- \\
  3/2 & 4 & 9 & 11 & 10 & 6 & 3 & 1 & --- & --- & --- & --- & --- & --- & --- & --- & --- & --- & --- & --- & --- & --- \\
  2 & 5 & 12 & 16 & 17 & 15 & 10 & 6 & 3 & 1 & --- & --- & --- & --- & --- & --- & --- & --- & --- & --- & --- & --- \\
  5/2 & 6 & 15 & 21 & 24 & 24 & 21 & 15 & 10 & 6 & 3 & 1 & --- & --- & --- & --- & --- & --- & --- & --- & --- & --- \\
  3 & 7 & 18 & 26 & 31 & 33 & 32 & 28 & 21 & 15 & 10 & 6 & 3 & 1 & --- & --- & --- & --- & --- & --- & --- & --- \\
  7/2   & 8 & 21 & 31 & 38 & 42 & 43 & 41 & 36 & 28 & 21 & 15 & 10 & 6 & 3 & 1 & --- & --- & --- & --- & --- & --- \\
  4 & 9 & 24 & 36 & 45 & 51 & 54 & 54 & 51 & 45 & 36 & 28 & 21 & 15 & 10 & 6 & 3 & 1 & --- & --- & --- & --- \\
  9/2 & 10 & 27 & 41 & 52 & 60 & 65 & 67 & 66 & 62 & 55 & 45 & 36 & 28 & 21 & 15 & 10 & 6 & 3 & 1 & --- & --- \\
  5 & 11 & 30 & 46 & 59 & 69 & 76 & 80 & 81 & 79 & 74 & 66 & 55 & 45 & 36 & 28 & 21 & 15 & 10 & 6 & 3 & 1 \\
  \hline
\end{tabular}
\end{table*}

For classical model weight factors can be calculated analytically (see Appendix \ref{app:density} for details), the corresponding expressions are:

\begin{equation}\label{eqn:D3}
    D_3(S)=
    \left\lbrace
      \begin{array}{c}
        {S^2}/{2},~~~0\leq S<1 \\
        {S(3-S)}/{4},~~~1\leq S\leq 3\\
      \end{array}
    \right.
\end{equation}

\begin{equation}\label{eqn:D4}
    D_4(S)=
    \left\lbrace
      \begin{array}{c}
        {S^2}\left(1-3S/8\right)/2,~~~0\leq S<2 \\
        S\left(1-{S}/{4}\right)^2,~~~2\leq S\leq 4\\
      \end{array}
    \right.
\end{equation}

\noindent Note that slope of $D_{3,4}(S)$ changes at $S=1,2$, correspondingly.

\section{Magnetization and differential susceptibility of the toy model: reminiscence of the plateau and its characterization}

\begin{figure}
\centering
\includegraphics[width=\figwidth]{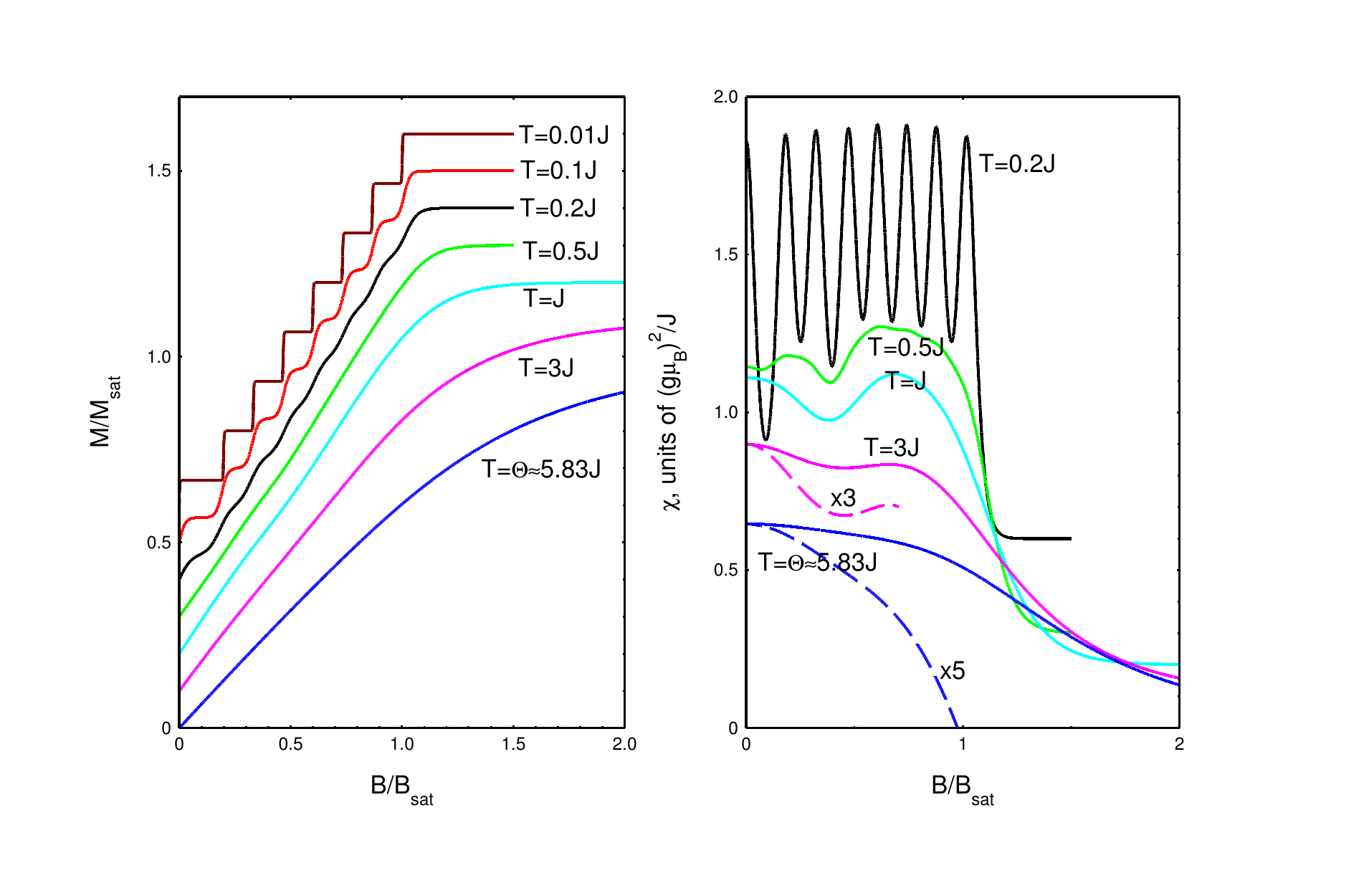} 
\caption{Left panel: magnetization curves for triangular spin cluster with $S=5/2$ at different temperatures, curves are shifted for better presentation. Right panel: differential susceptibility field dependencies for triangular spin cluster with $S=5/2$ at different temperatures, curves are shifted for better presentation. Dashed curves at the left panel are y-magnified with the factor shown. }
\label{fig:m_and_chi_quantum}
\end{figure}

\begin{figure}
  % Requires \usepackage{epsfig}
  \centering
  \includegraphics[width=\figwidth]{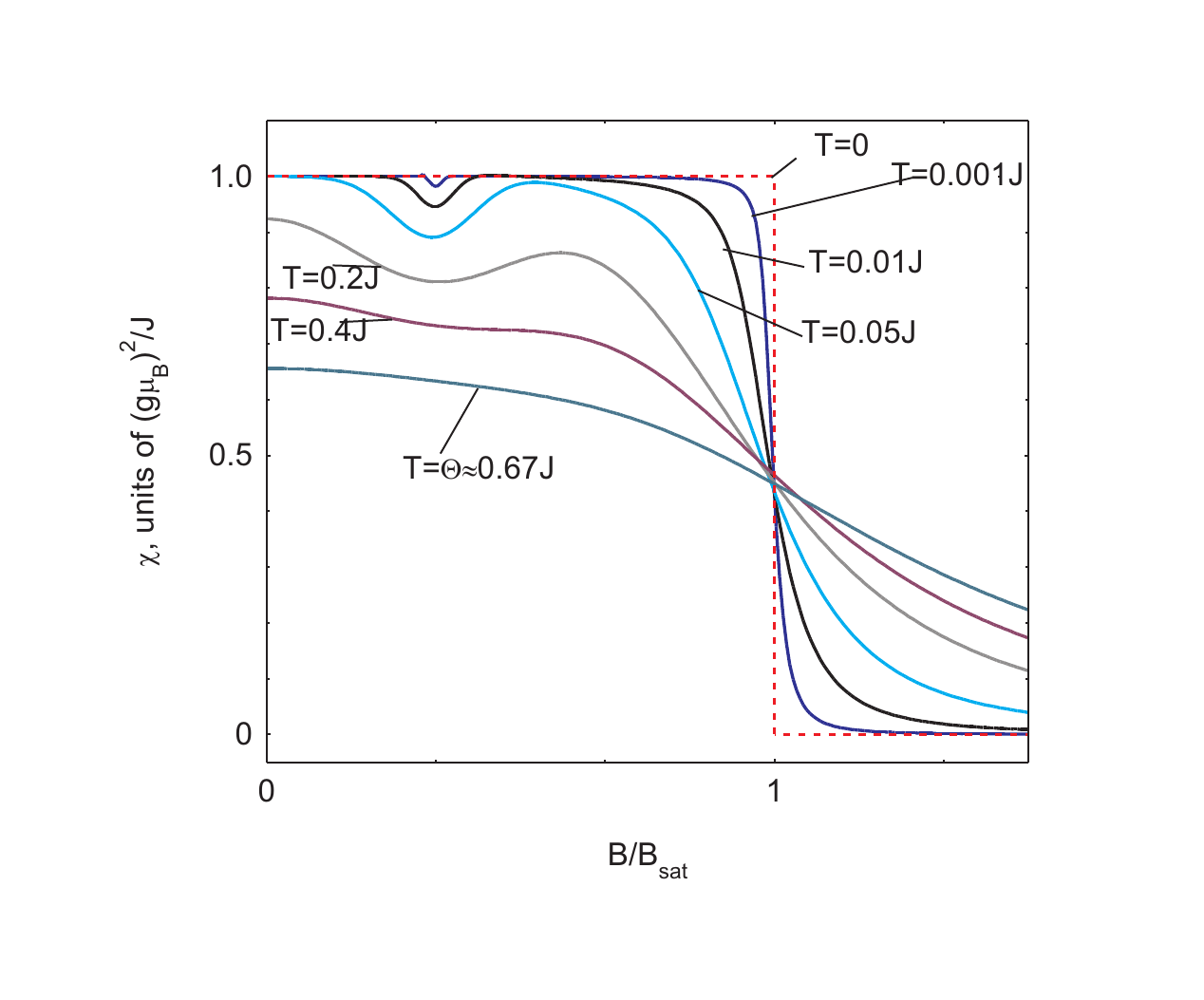}
  \caption{Differential susceptibility curves for triangular cluster of classical unit spins at different temperatures.}\label{fig:chi-classic}
\end{figure}

\begin{figure}
\centering
\includegraphics[width=\figwidth]{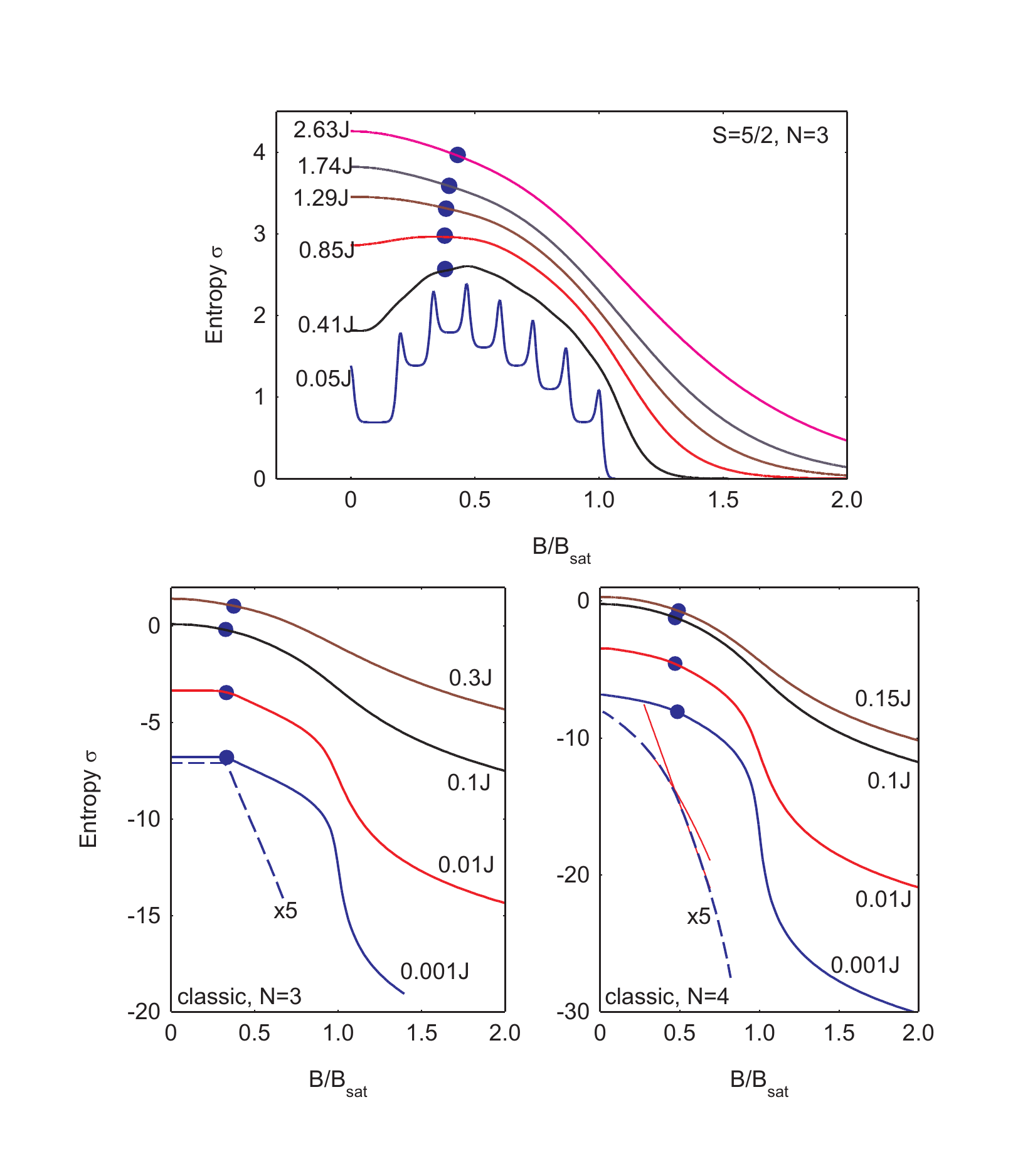}
\caption{Entropy field dependence at different temperatures for triangular cluster of quantum spins $S=5/2$ (top) and trigonal
and tetrahedral clusters of classical spins (bottom). Filled circles shows position of minimum of differential susceptibility at corresponding temperature.
Dashed curves (for classical models) are 5-fold Y-expanded  to make $\sigma(B)$ slope change more pronounced, thin red lines
on bottom right panel are smooth guides to the eye marking this slope change. Temperature values at the top panel equal
in units of $J S(S+1)$ to 0.0057, 0.0463, 0.097, 0.148, 0.199 and 0.3, correspondingly.}
\label{fig:entropy}
\end{figure}

\begin{figure}
\centering
\includegraphics[width=\figwidth]{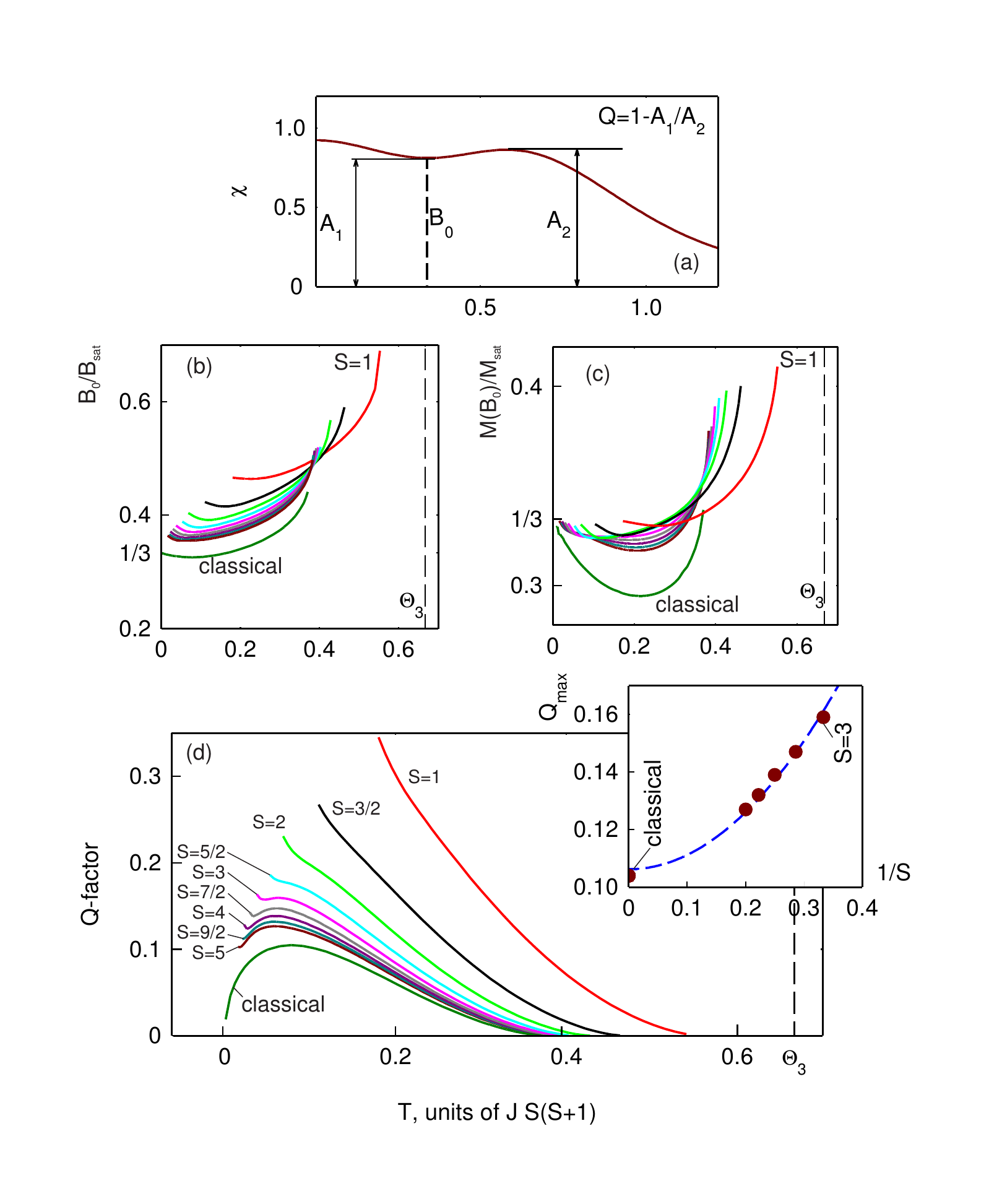}
\caption{(a) Scheme of quantitative characterization of plateau-like feature of toy model magnetization process. Results for triangular ($N=3$) clusters for $S\geq1$ (classical limit including): (b) Temperature dependence of differential susceptibility minimum position. (c) Temperature dependence of the magnetization at susceptibility minimum. (d)  Temperature dependence of the plateau quality factor, inset: dependence of maximal plateau quality factor on spin in the triangle vertice for $S\geq3$, dashed curve is guide to the eye. Color scheme on panels (b), (c), (d) is the same. Vertical dashed line on panels (b), (c), (d) is a Curie-Weiss temperature, temperature for quantum model is expressed in the units of $J S(S+1)$, temperature for classical model is expressed in the units of $J$.}
\label{fig:characterization3}
\end{figure}

\begin{figure}
\centering
\includegraphics[width=\figwidth]{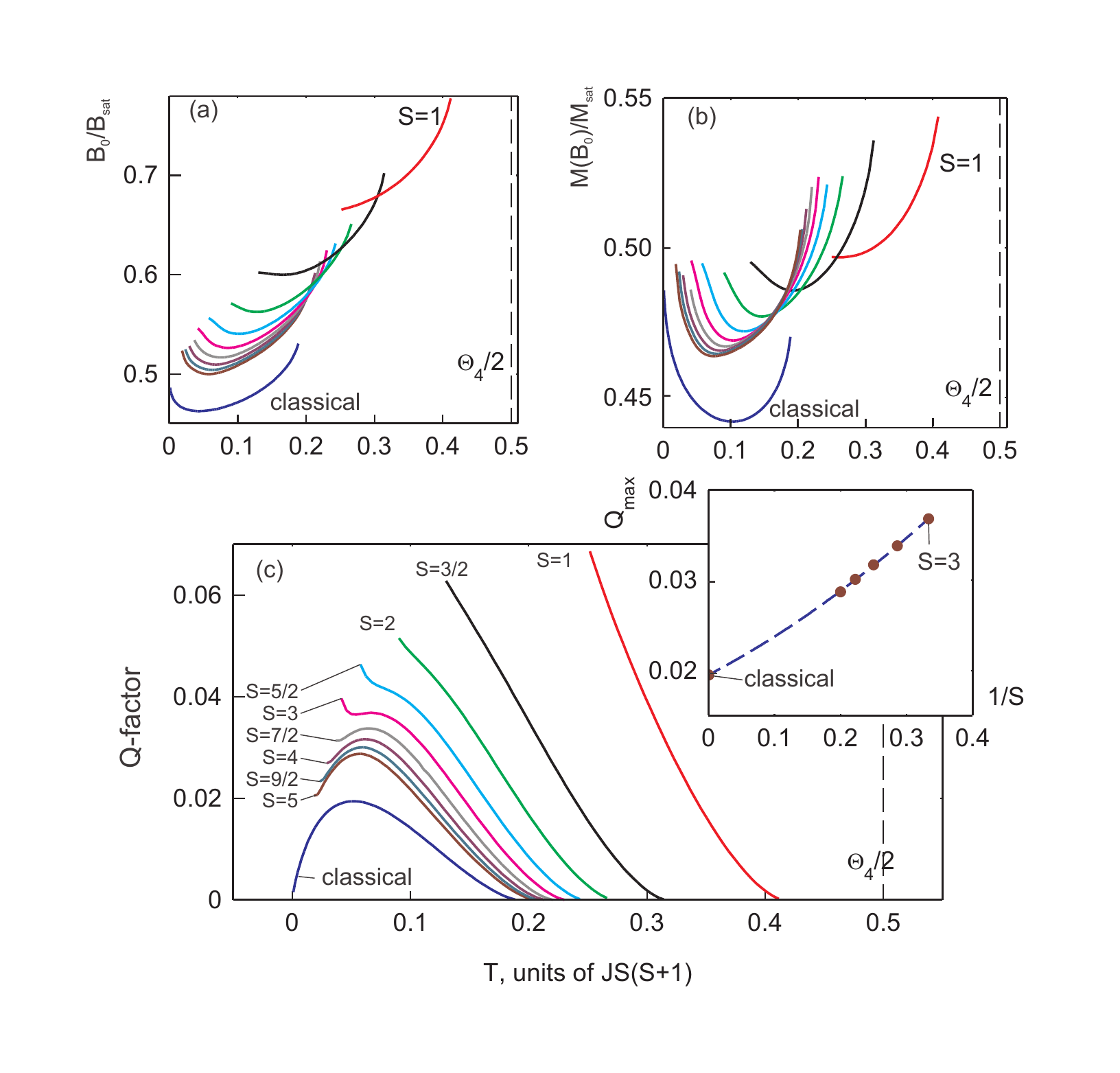}
\caption{Quantitative characterization of plateau-like feature of toy model magnetization process, results for tetrahedral ($N=4$) clusters for $S\geq1$ (classical limit including): (a) Temperature dependence of differential susceptibility minimum position. (b) Temperature dependence of the magnetization at susceptibility minimum. (c) Temperature dependence of the plateau quality-factor, inset: dependence of maximal plateau quality factor on vertice spin  for $S\geq3$, dashed curve is guide to the eye. Color scheme is the same for all panels. Vertical dashed line on panels (a), (b), (c) is one-half of the  Curie-Weiss temperature, temperature for quantum model is expressed in the units of $J S(S+1)$, temperature for classical model is expressed in the units of $J$.}
\label{fig:characterization4}
\end{figure}

First, we recall that at $T=0$ magnetization of a quantum model rises stepwise as higher spin states in turn became a ground state. Maximal polarization is reached at zero-temperature saturation field $B_{sat}=\frac{N J S}{g\mu_B}$, magnetization steps are separated by $\Delta B=\frac{J}{g\mu_B}$. The steps of magnetization became smeared as temperature rises. High-temperature limit can be accessed via standard high-temperature series, these calculations show that high-temperature zero-field susceptibility follows Curie-Weiss law
\begin{equation}\label{eqn:CW}
  \chi=\frac{N(g\mu_B)^2 S(S+1)}{3} \frac{1}{T+\Theta}
\end{equation}

\noindent
here Curie-Weiss temperature equals $\Theta_3=\frac{2}{3} J S(S+1)$ for triangular cluster ($N=3$) and $\Theta_4=J S(S+1)$ for tetrahedral cluster ($N=4$).\footnote{However, numerical analysis shows that the ``local'' Curie-Weiss temperature $\Theta=(g\mu_B)^2 S(S+1)/(3\chi)-T$ approaches its high temperature limit within 1\% only at $T>50 J S(S+1)$.} At low temperatures zero-field susceptibility diverges as $1/T$ for triangular clusters with half-integer spins ($S=1/2$ ground state) and exponentially approaches zero for triangular clusters with integer spin and all tetrahedral clusters ($S=0$ ground state).

For the classical model saturation field is $B_{sat}=\frac{N J}{g\mu_B}$, magnetization process at zero temperature is linear up to saturation field with susceptibility $\chi_0=(g\mu_B)^2/J$. High-temperature susceptibility of the classical model follows Curie-Weiss law with Curie-Weiss temperatures $\Theta_3=\frac{2}{3}J$ and $\Theta_4=J$ for triangular ($N=3$) and tetrahedral ($N=4$) clusters, correspondingly.

Figure \ref{fig:m_and_chi_quantum} shows modeled magnetization and differential susceptibility curves for trigonal cluster of spins $S=5/2$ at different temperatures. As expected, step-like increase of magnetization is smeared with the temperature and almost vanishes at $T\simeq J$ and high temperatures magnetization curves evolve toward standard Brillouine curve of a paramagnet. However, differential susceptibility tales another story: as expected, its field dependence is a series of sharp peaks at low temperature, these peaks broadens on heating and form sort of ``wobbling'' at $T\simeq J$. However, as this ``wobbling'' disappears on further heating, {\em local minimum} of $\chi(B)$ remains at approximately 1/3 of the saturation field. This local minimum remains visible to the temperatures of the order of Curie-Weiss temperature, which is much higher than $J$ for large spins. For the classical model (see Figure \ref{fig:chi-classic}) differential susceptibility is constant up to saturation field at $T=0$, however local minimum of differential susceptibility appears on heating and, again, survives up to the temperatures of the order of Curie-Weiss temperature.

Similar effect, so of smaller amplitude, was observed for tetrahedral clusters as well. In the case of $N=4$ the local minimum of differential susceptibility is located at approximately half of the saturation field.

As it was already mentioned in the Introduction, magnetization plateau at 1/3 and 1/2 of the magnetization saturation value are predicted and observed for antiferromagnets on triangular (and kagom\'{e}) lattice and on the pyrochlore lattice, correspondingly. As these plateaus frequently appears at the high-field region, they are sometimes detected in pulsed field experiments and differential susceptibility is the experimentally measured quantity there \cite{rbfeplateau}. The magnetization plateaus appears then as minima of differential susceptibility. Ideally, differential susceptibility drops to zero at plateau range, but experimentally it is known to remain finite \cite{rbfeplateau}. Thus, the local minimum of differential susceptibility of our toy model strongly resembles this experimental situation. It looks like some reminiscence of the magnetization plateau of macroscopic magnet is ``hidden''  within the properties of its building block. However, magnetization plateau of a macroscopic magnet is a distinct phase and plateau edges are marked by real phase transition, while for microscopic toy model differential susceptibility evolves smoothly.

This local  minimum of magnetization for our toy model is, essentially, a classical effect -- it survives transition to the classical limit, it is due to maximal weight factor for the states with intermediate total spin. The reason for the observed decrease of differential susceptibility is statistical and it closely resembles ``order by disorder'' mechanism stabilizing plateau phase in macroscopic case. As weight factor $D(S_{tot})$ demonstrates maximum at some intermediate spin value $S_0$ low-temperature entropy $\sigma(T,B)$ could have a maximum as a  function of applied magnetic field close to the field at which $S=S_0$ spin multiplet becomes a ground state of the cluster. Since free energy is $F=E-T\sigma$ (here $\sigma$ is an entropy) and magnetization is

\begin{equation}\label{eqn:mag-and-S}
    \vect{M}=-\frac{\partial F}{\partial \vect{B}}=-\frac{\partial E}{\partial \vect{B}}+T\frac{\partial \sigma}{\partial \vect{B}},
\end{equation}

\noindent then magnetization tends to increase faster on approaching entropy maximum (${\partial \sigma}/{\partial {B}}>0$) and to decrease somehow after the entropy maximum. This corresponds to the leveling of $M(B)$ curve and to the local minimum of differential susceptibility. These qualitative arguments are in agreement with numerical calculations for cluster of quantum spins (see upper panel of Figure \ref{fig:entropy}). For the classical models (lower panels of Figure \ref{fig:entropy}) entropy maximum is absent and position of susceptibility minimum at low temperatures is marked by sharp change of the slope of $\sigma(B)$ curve: at this point entropy starts to decrease faster with the applied field which, again, acts towards leveling of $M(B)$ curve. Taking leading exponent in the Eqn.(\ref{eqn:Z-classic}) at low temperatures $T\ll J$ and below saturation field one can obtain for $Z$ and entropy of classical model:

\begin{eqnarray}
Z&\approx& \left(2\pi \frac{T}{J}\right)^{3/2} \frac{D_N(g\mu_B B/J)}{(g\mu_B B/J)^2} e^{(g\mu_B B)^2/(2TJ)}\\
\sigma&\approx&const+\frac{3}{2} \ln \frac{T}{J}+\ln \frac{D_N(g\mu_B B/J)}{(g\mu_B B/J)^2}
\end{eqnarray}

\noindent recalling Eqns.(\ref{eqn:D3}) and (\ref{eqn:D4}) we can see that, indeed, entropy maximum is absent at low temperatures, and, since $D_3(S)\propto S^2$ for low $S$, low-field entropy of trigonal cluster is field-independent. Strong field dependence of the low-temperature entropy in the vicinity of the saturation field gives rise to the enhanced magnetocaloric effect, which is a known effect both for molecular magnets \cite{schnack}, model cubeoctahedron system \cite{cubeok} and for the related macroscopic frustrated pyrochlore-lattice antiferromagnet \cite{zhitomirskii,zhitsosin}.

To quantify this plateau-like feature of our toy model we computed following quantities (see also scheme at the Figure \ref{fig:characterization3}-a) for different spin values and for classical model: position of the local minimum of differential susceptibility $B_0$, magnetization value at $B_0$ and quality factor of the plateau defined as

\begin{equation}\label{eqn:Q-factordef}
  Q=1-\frac{\chi(B_0)}{\max_{B>B_0}\chi(B)}.
\end{equation}
\noindent Here $\max_{B>B_0}\chi(B)$ is the maximal value of differential susceptibility to the right from $B_0$, quality factor is equal to unity for ideal plateau (with $\chi(B_0)=0$) and turns to zero as plateau-like feature disappears. On cooling magnetization process of the quantum model evolves towards step-like behavior accompanied by the ``wobbling'' of the differential susceptibility. As amplitude of these oscillations increase local minimum became ill defined and any numeric procedure will be locked to the minimum of susceptibility between two nearest steps, which has completely different origin.  To exclude this effect we cut the data on low temperature side at some temperature where regular oscillations of differential susceptibility became remarkable. The results are shown on the Figures \ref{fig:characterization3}, \ref{fig:characterization4}. Extreme quantum case of $S=1/2$ is not shown on these Figures, as it evolves in a more trivial way: at low temperature magnetization process of triangular cluster has two steps at zero field (corresponding to polarization of two-fold degenerated $S=1/2$ ground state) and at saturation field (corresponding to the transition to fully polarized $S=3/2$ state), and that of tetrahedral cluster shows two steps at half of the saturation field ($S=0$ to $S=1$ transition) and at the saturation field (transition to fully polarized $S=2$ state). The local minima of differential susceptibility are locked between these steps and remains there on heating. We will demonstrate data for $S=1/2$ in an Appendix \ref{app:onehalf} for completeness.

As it was already mentioned, plateau-like feature is observed (see Figures \ref{fig:characterization3},\ref{fig:characterization4}) in an extended temperature range starting for quantum models from $T\sim J$ up to $T\sim 0.5 J S (S+1)$ and for the classical models from $T=0$ to $T\sim 0.2...0.3J$. This indicates presence of two temperature scales for the magnetization process of our clusters: large scale $T_1\sim\Theta\sim J S (S+1)$ corresponds to the total spread of the cluster sublevels in the energy domain, while small one$T_2\sim J$ corresponds to the difference between nearest sublevels. At $T>T_1$ all sublevels of the cluster are populated and spins in the cluster vertices can be considered being almost independent. At $T<T_2$ only lowest levels are populated and cluster magnetization is determined by total spin of its ground state.  For large $S$ these temperature scales could differ by the order of magnitude resulting in intermediate regime where effects of discreteness of energy sublevels are already smeared out by temperature fluctuations, but spin-spin correlations remains important.
One can see that plateau-like local minima of differential susceptibility are observed around 1/3 and 1/2 of the saturation field and magnetization for triangular and tetrahedral clusters, correspondingly. Plateau-like feature is better developed for the triangular cluster, maximal value of the  quality factor for the classical model is about 0.10 for trigonal cluster and 0.02 for tetrahedral cluster. On heating quality factor turns to zero at certain temperature: at this point minimum of the differential susceptibility disappears.

\section{Comparison with the experiment on RbFe(MO$_4$)$_2$}
\begin{figure}
  % Requires \usepackage{epsfig}
  \centering
  \includegraphics[width=\figwidth]{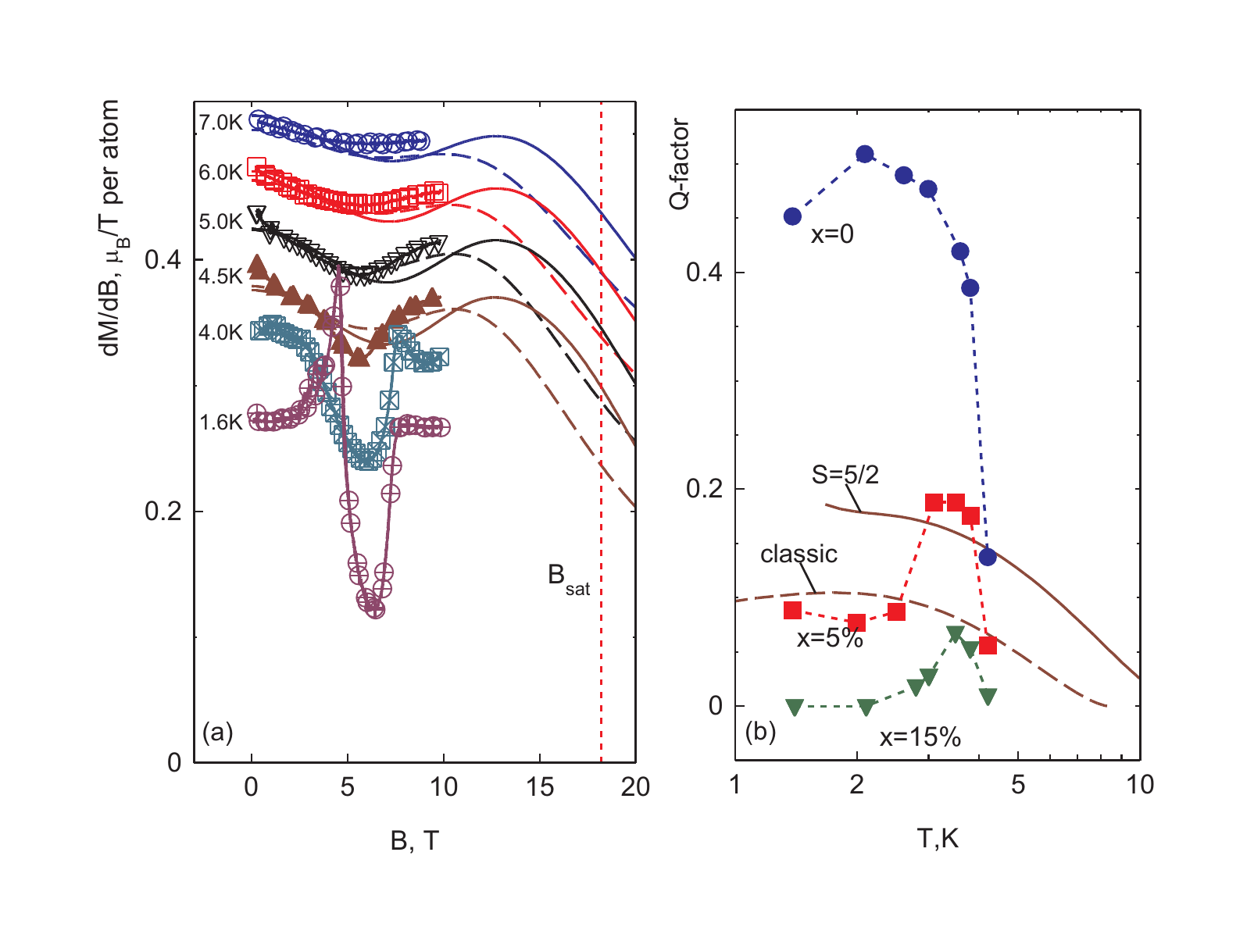}
  \caption{(a) Comparison of experimentally measured differential susceptibility of the triangular antiferromagnet RbFe(MO$_4$)$_2$ (lines with symbols, from Ref.\cite{rbfesmirnov}) with the toy model calculations (solid lines for quantum model, dashed lines for classical model). Zero-field Neel temperature for RbFe(MO$_4$)$_2$  is about 4.3K. Curves for different temperatures are Y-offset for better presentation, model parameters are tuned to reproduce low  temperature saturation field and magnetization as described in the text. (b) Comparison of experimentally measured plateau quality factor for Rb$_{1-x}$K$_x$Fe(MO$_4$)$_2$ (lines with symbols, from Ref.\cite{rbfediluted}) with toy model calculations (solid lines for quantum model, dashed lines for classical model). Model parameters are tuned as described in the text. }\label{fig:expdata}
\end{figure}

Triangular lattice antiferromagnet RbFe(MO$_4$)$_2$  is one of the model systems for magnetization plateau demonstration.\cite{rbfeplateau,rbfesmirnov} Its measured phase diagram was found to scale very well with the theoretical predictions. However, accurate measurements of magnetization \cite{rbfesmirnov} showed that minimum of differential susceptibility exists even in the paramagnetic phase above the Neel point. Recently, magnetization of a  magnet  Rb$_{1-x}$K$_x$Fe(MO$_4$)$_2$ with  chaotic modulation of the exchange bonds was studied as well, \cite{rbfediluted} these studies demonstrated that macroscopic collinear phase can be suppressed by substituting  approx. 15\% of Rb by K at low temperatures, but on heating collinear phase could be reestablished by thermal fluctuations. Indeed, small minimum of differential susceptibility was observed,\cite{rbfediluted} temperature dependence of Q-factor  of this minimum is qualitatively close to our model prediction: it starts from zero at $T=0$, demonstrates maximum at some temperature and then again turns to zero.

Assuming that this behavior of the magnetization in paramagnetic phase or in the diluted magnet is due to short-range correlations, we compare now measured magnetization curves Q-factor with the results of our toy model: triangular cluster is a natural example of ``shortly correlated'' spins on a triangular lattice. When performing this comparison one have to keep in mind essential difference of macroscopic magnet and microscopic cluster: magnetization plateau of a macroscopic magnet, even if $Q<1$ for some reason, is a distinct phase with clear phase boundaries, it differs from other phases by symmetry. In the case of model microscopic cluster its state at the minimum of differential susceptibility is not specially distinguished. However, as we will show below, some features of the toy model are really close to the real macroscopic system.

To scale model parameters to that of real magnet we will use saturation field which is 18.2 T at 1.3...1.5 K in RbFe(MO$_4$)$_2$ and does not change remarkably with Rb/K substitution. Zero-field Neel temperature for RbFe(MO$_4$)$_2$ is 4.2K, it increases with the field reaching approx. 4.5K at 6T. The $g$-factor value is known from ESR spectroscopy\cite{rbfesmirnov} and is equal to 2.2, spin of the magnetic Fe ions is $S=5/2$. This yields exchange integral for the quantum model $J=\frac{g \mu_B B_{sat}}{N S}\approx 3.6 K$. Classical model assumes $|\vect{S}=1|$, thus we have to tune effective $g$-factor to fit saturation magnetization value, which is $g\mu_B$, hence $g_{cl}=2.2\times 5/2=5.5$. Then the effective exchange coupling constant of our model will be $J=\frac{g_{cl} \mu_B B_{sat}}{N}\approx 22.4 K$. The model Curie-Weiss temperatures are then 21K for $S=5/2$ quantum model and 15K for classical model, correspondingly.

These constants fixed, we can calculate differential susceptibility per ion and Q-factor  without other tunable parameters, comparison of the modeled curves with experiment is shown at the Figure \ref{fig:expdata}. This comparison shows qualitative similarity of model and experiment, note also that amplitude of modeled differential susceptibility is calculated without any tunable scaling factor. Naturally, microscopic toy model can not reproduce any features related to the phase transitions of a macroscopic system: sharp edges of the plateau phase at $T<4.2$K and Q-factor in the ordered phase of pure RbFe(MO$_4$)$_2$ differs very strongly from model. As for the case of magnet with chaotic modulation of exchange bonds Rb$_{1-x}$K$_x$Fe(MO$_4$)$_2$, maximal value of the Q-factor is close to the model calculations for large $x$. However, again, Q-factor in Rb$_{1-x}$K$_x$Fe(MO$_4$)$_2$ starts to rise from zero at certain temperature, which is the temperature of phase transition at which $uud$ phase revives due to increasing thermal fluctuations, while in case of microscopic cluster Q-factor continuously decreases down to $T=0$.

\section{Stability of differential magnetization close to the saturation field}
\begin{figure}
\centering
\includegraphics[width=\figwidth]{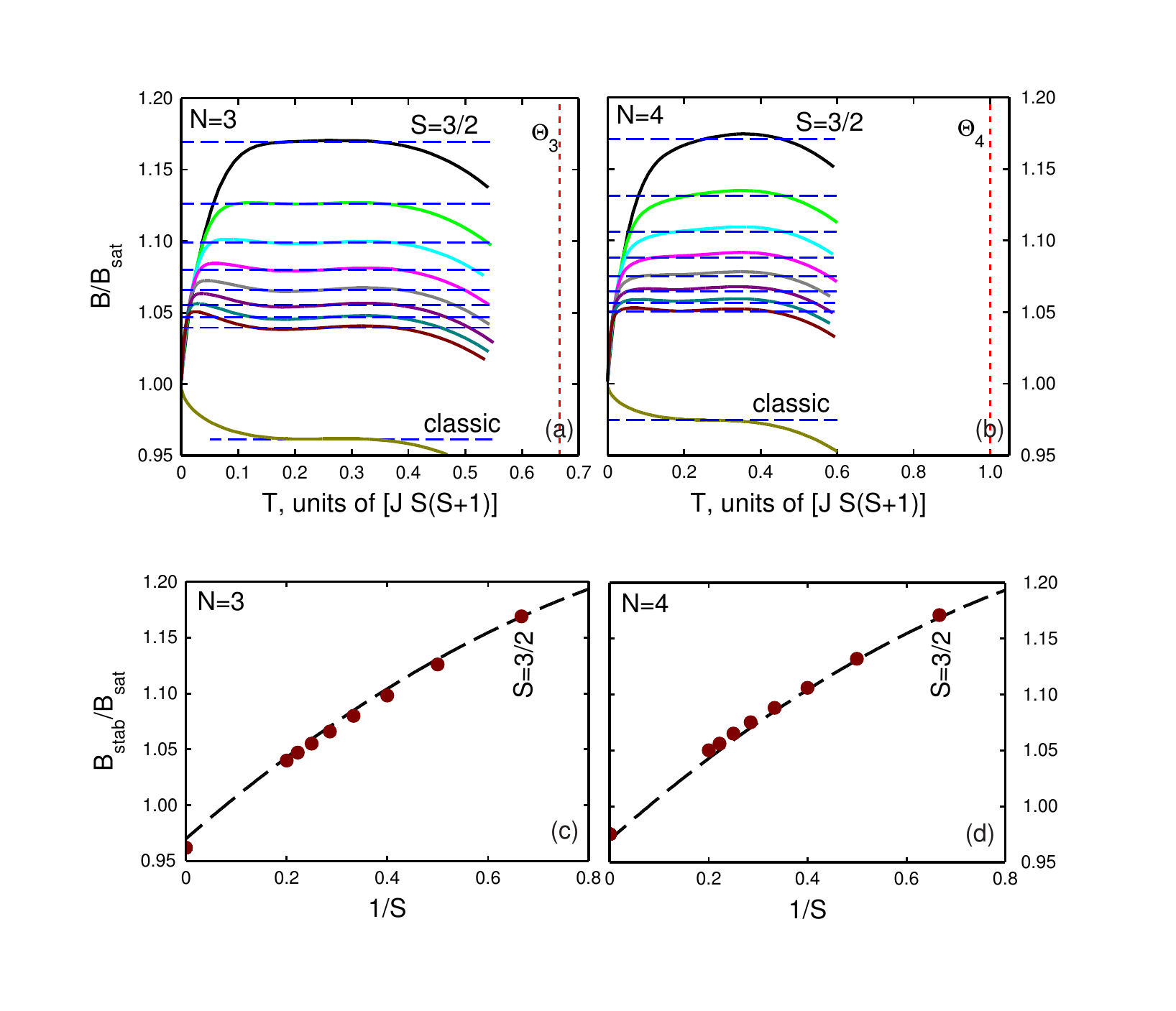}
\caption{(a) and (b) Temperature dependences of magnetic field at which differential susceptibility is equal to $\frac{1}{2}\frac{(g\mu_B)^2}{J}$ for different spin at the cluster vertice ($3/2 \leq S\leq 5$ and classical model) for trigonal ($N=3$) and tetrahedral ($N=4$) clusters, correspondingly. Horizontal dashed lines are fit of the stability ranges, vertical dashed lines mark Curie-Weiss temperature, temperature for quantum model is expressed in the units of $J S(S+1)$, temperature for classical model is expressed in the units of $J$. (c) and (d) Value of magnetic field at the stability range as a function of inverse spin for trigonal ($N=3$) and tetrahedral ($N=4$) clusters, correspondingly. Dashed line is empirical fit $B/B_{sat}=0.97+0.39/S-0.14/S^2$. }
\label{fig:stabpoint}
\end{figure}

Finally, we briefly comment on the high-field behavior of differential susceptibility. At $B\gg B_{sat}$ $\chi$ should approach zero as cluster magnetization approaches saturation. Characteristic scale of the susceptibility is given by its $T=0$ value for classical model  $\chi_0=\frac{(g\mu_B)^2}{J}$.

We have found that close to the saturation field the value of magnetic field, at which differential susceptibility equals to $\chi_0/2$, is almost temperature independent (changes within 1\%) at a quite extended temperature range (Figure \ref{fig:stabpoint}). This observation can be of use as a tool to determine exchange constant of the cluster from the magnetization data. This stable behavior of differential susceptibility is observed for the spins $S\geq 3/2$ (classical limit including) at the temperatures from approx. $0.1 J S (S+1)$ to  approx. $0.4 J S (S+1)$ (from approx. $0.1 J $ to  approx. $0.4 J$ for classical model) for trigonal and tetrahedral clusters alike, the magnetic field in this stability range depends on the value of the spin in the cluster vertice. This dependence can be empirically fitted for both trigonal and tetrahedral clusters as $$\frac{B_{stab}}{B_{sat}} =0.97+0.39\frac{1}{S}-0.14\frac{1}{S^2}.$$ Interestingly, the $\chi=\chi_0/2$ point shifts to fields above $B_{sat}$ for quantum models and to fields below $B_{sat}$ for classical model.

\section{Conclusions.}

Magnetization process of the triangular and tetrahedral cluster has two important energy scales: exchange coupling parameter and Curie-Weiss temperature. These energy scales differ by a factor of $S(S+1)$ for quantum model. At lowest temperatures $T\ll J$ magnetization curve of a cluster of quantum spins is step-like, at high temperatures $T\gg \Theta$ susceptibility of the cluster follows Curie-Weiss law. However, in the intermediate temperature range $J\lesssim T\lesssim\Theta$ strong correlations results in nonlinearity of magnetization process accompanied by a local minimum of differential susceptibility (magnetization curve slope) at approximately the same field (and at approximately the same magnetization level) where magnetization plateau is observed for a triangular or pyrochlore-lattice antiferromagnets.

We have analyzed this effect on a set of quantum models with $1/2\leq S\leq 5$ and on a classical model. The positions, quality factors and ranges of existence for this plateau-like  feature are determined. Model results are compared to the known experimental facts. We have demonstrated that magnetization of toy model clusters does bear some analogies to the magnetization process of macroscopic triangular lattice magnet at paramagnetic phase or in the presence of strong disorder. However, microscopic toy model and macroscopic system became qualitatively different in the ordered phase as phase transitions appears in macroscopic magnet.

\section*{Acknowledgements}

Author thanks Prof.L.E.Svistov (Kapitza Institute) for drawing attention to this problem and Prof.A.I.Smirnov  (Kapitza Institute) for providing experimental data for comparison and useful discussions.

The work was supported by Russian Science Foundation Grant No.17-02-01505, Author work at HSE was supported by Program of fundamental studies of Higher School of Economics.

\appendix

\section{Details of weight factor calculations \label{app:density}}

We calculate the weight factor for quantum model by simple ``brute force'' counting.
For finite quantum mechanical model one can easily count all possible spin projections, which runs from
-N S to N S . Each of these projections can be constructed in several ways. Total spin of the
cluster can take values from $N S$ to 0 or 1/2. To calculate weight factor one simply has to
note that if some nonnegative spin projection $S_z$ can be constructed by $n_{S_z}$ ways and higher
projection $(S_z+1)$ can be constructed by $n_{S_z+1}$ ways, then weight factor for total spin $S=S_z$ is exactly $D_N(S)=n_{S_z}-n_{S_z+1}$ , as
all total spin states that have the spin projection $(S_z+1)$ have to include spin projection $S_z$ in their spin multiplet as
well. Results are summed up in the Tables \ref{tab:D3even}, \ref{tab:D3odd} and \ref{tab:D4}. We do not obtained compact combinatorial expressions for the weight factors, however we can note that for $N=3$ found values correspond to $D(S_{tot})=1+2 S_{tot}$ for $S_{tot}\leq S$ and to $D(S_{tot})=1+3S-S_{tot}$ for $S_{tot}>S$. It is also interesting to note here that weight factor for the singlet ($S=0$) state of tetrahedral cluster increases with increasing spin as $(2S+1)$, being a precursor of the macroscopic degeneracy of pyrochlore lattice antiferromagnets.

For classical model we will start from the case $N=2$. We are interested in total spin of the pair, which does not change on simultaneous rotation of both spins. Thus, we can fix one spin and weight factor can be calculated from all possible orientations of the second spin with respect to the first. If $\Theta$ is a polar angle describing second spin direction selected in such a way that $\Theta=0$ corresponds to antiparallel orientations of spins, then total spin of the pair is $$S^2=2(1+cos\Theta)$$ (spin vectors of the classical model are unit vectors). Then we differentiate this equation $$S d S=-sin\Theta d\Theta$$ and compare it with the fraction of realization of spin configurations within the same angle interval $$\frac {d n}{n}=\frac{2\pi sin\Theta d\Theta}{4\pi}.$$ This gives directly the weight factor

\begin{equation}\label{eqn:app-D2}
D_2(S)=\frac{1}{n} \frac{d n}{d S}=\frac{S}{2}
\end{equation}
\noindent
here $0\leq S\leq 2$.

The result could look counterintuitive: the probability of $S=0$ configuration appears to be much less then the probability of $S=2$ configuration. This is actually due to the fact that for antiparallel orientation ($S=0$) small deviations from the exact antiparallel orientation results in the appearance of the total spin {\em linear} in deviation angle, while for parallel orientation ($S=2$) small deviations lead to decrease of the total spin {\em quadratic} in deviation angle. Also one can note on this simple example that classical and quantum weight factors differ strongly here: in the quantum case for the pair of equal spins each value of the total spin from zero to $2 S$ is unique.

For the case of three spins $N=3$ we, again, can fix the direction of one spin $\vect{S}_1$. Then we sum up two remaining spins $\vect{\sigma}=\vect{S}_2+\vect{S}_3$, the length distribution for spin $\vect{\sigma}$ is found above. Total spin is equal to $$S^2=1+\sigma^2+2\sigma cos\Theta,$$ here polar angle $\Theta$ is selected as above.

All possible configurations are confined in the plane $(\sigma,cos\Theta)$ with $0\leq\sigma\leq 2$ and $-1\leq cos\Theta\leq 1$. Fraction of realization of spin configurations within element $d\sigma d cos\Theta$ is $$\frac{d n}{n}= D_2(\sigma) d\sigma d(cos\Theta)/2=\frac{\sigma}{4}d\sigma d(cos\Theta).$$ From isoline equation $cos\Theta=(S^2-1-\sigma^2)/(2\sigma)$ one can derive $d(cos\Theta)=S dS/\sigma$. This results in differential weight factor $$\frac{1}{n}\frac{d n}{d S}=\frac{S}{4}d\sigma,$$ which have to be integrated over possible $\sigma$.

Possible $\sigma$ range depends on $S$ value:  $(1-S)\leq\sigma\leq(1+S)$ for $S<1$, and $(S-1)\leq\sigma\leq 2$ for $S>1$. This yields

\begin{equation}\label{eqn:app-D3}
        D_3(S)=
    \left\lbrace
      \begin{array}{c}
        \frac{S^2}{2},~~~0\leq S<1 \\
        \frac{S(3-S)}{4},~~~1\leq S\leq 3\\
      \end{array}
    \right.
\end{equation}

Finally, for the case of tetrahedral cluster $N=4$ we follow the same route. We fix spin vector $\vect{S}_1$ and sum up all other vectors to the spin vector $\vect{\sigma}=\vect{S}_2+\vect{S}_3+\vect{S}_4$ with known length distribution and derive differential weight factor
$$\frac{1}{n}\frac{d n}{d S}=\frac{S}{2} \frac{D_3(\sigma)}{\sigma} d\sigma,$$
which have to be integrated over possible $\sigma$. Possible $\sigma$ ranges are: $(1-S)\leq\sigma\leq(1+S)$ for $S<1$ , and   $(S-1)\leq\sigma\leq(1+S)$  for $1<S<2$, and $(S-1)\leq\sigma\leq 3$ for $S>2$. This yields

\begin{equation}\label{eqn:app-D4}
    D_4(S)=
    \left\lbrace
      \begin{array}{c}
        \frac{S^2}{2}\left(1-\frac{3}{8}S\right),~~~0\leq S<2 \\
        S\left(1-\frac{S}{4}\right)^2,~~~2\leq S\leq 4\\
      \end{array}
    \right.
\end{equation}

Note, that $D_3(S)$ and $D_4(S)$ are equal to zero at extreme spin values (0 and $N$) and reach maximal values at $S=3/2$ and $S=16/9$ for triangular and tetrahedral clusters, correspondingly. The weight factors are continuous functions of total spin, however their derivative are discontinuous: slope of $D_N(S)$ changes at $S=1$ and $S=2$ for trigonal and tetrahedral clusters, correspondingly.

\section{Susceptibility curves for the extreme quantum case of $S=1/2$ \label{app:onehalf}}

Cases of $S=1/2$ spin clusters differ  from the higher spin cases analyzed above. The reason is that for small $S$ energy scales of exchange coupling $J$ and Curie-Weiss temperature $J S (S+1)$ do not differ strongly and, hence, the intermediate temperatures regime is simply absent.

\begin{figure}[h]
\centering
\includegraphics[width=\figwidth]{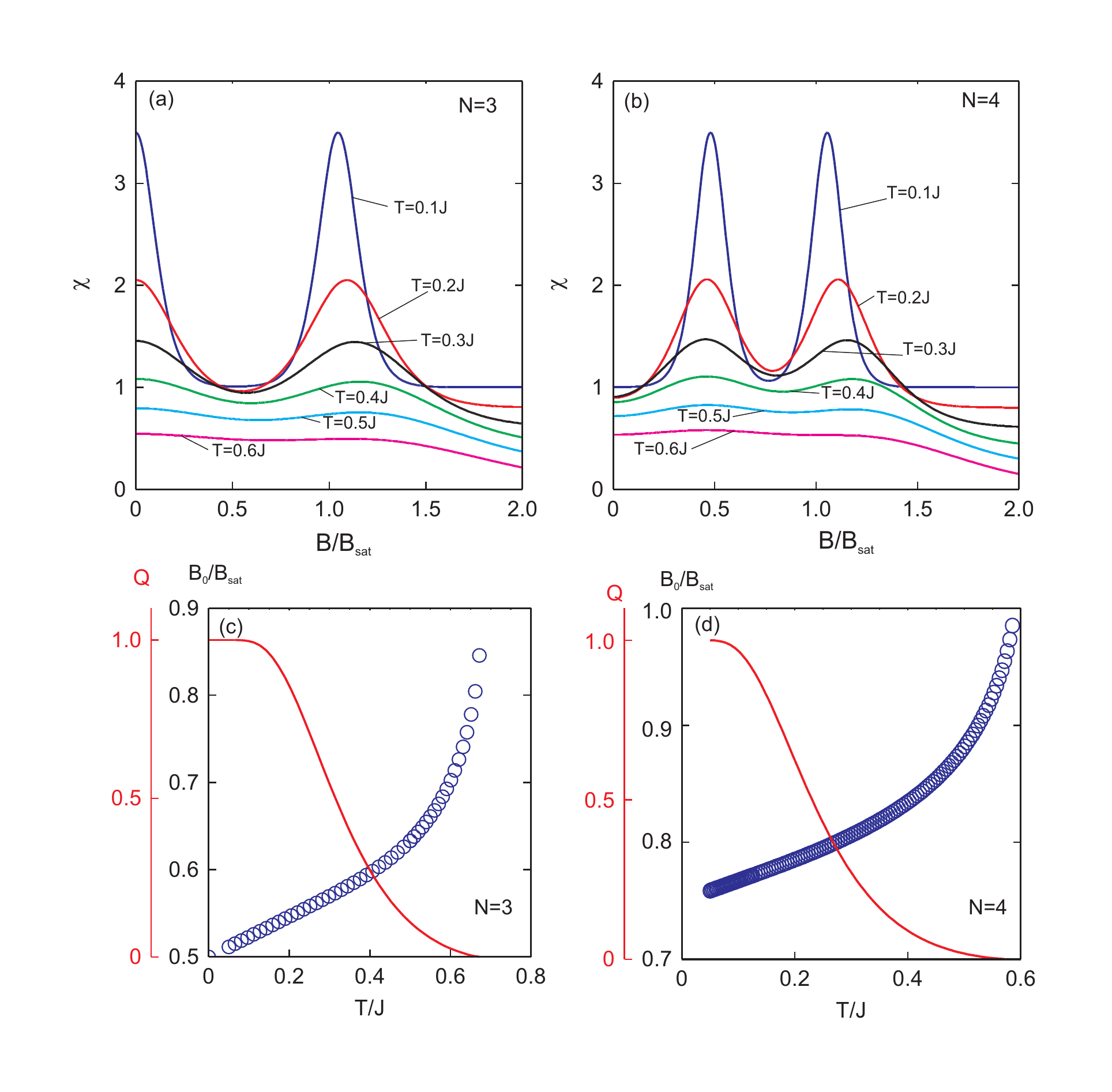}
\caption{(a) and (b) Examples of differential susceptibility curves for $S=1/2$ triangular (left) and tetrahedral (right) clusters. Curves are shifted for better presentation. (c) and (d) Temperature dependences of susceptibility minimum position (symbols) and its Q-factor (solid line) for $S=1/2$ triangular (left) and tetrahedral (right) clusters.}
\label{fig:extreem}
\end{figure}

 Modeling of magnetization and susceptibility curves (see Figure \ref{fig:extreem}) shows expected smearing of the magnetization steps on heating at $T\simeq 0.3 J$, differential susceptibility demonstrates  local minimum  up to the temperatures about $0.6...0.7 J$. On cooling this local minimum smoothly evolves toward position right between the magnetization steps: to $B_0=B_{sat}/2$ for the trigonal cluster and to $B_0=3 B_{sat}/4$ for the tetrahedral cluster. Consequently, plateau quality factor reaches unity at $T=0$. Similar locking of the susceptibility minimum between magnetization steps  also occurs for higher spins, however for higher spin temperature evolution of plateau-like feature parameters has demonstrated some sort of change (crossover-like) at locking: quality factor was starting to increase more rapidly, slope of the $B_0(T)$ curve was changing.

\end{document}